\documentclass[prc,twocolumn,showpacs]{revtex4}
\usepackage{natbib}
\usepackage{graphicx}
\usepackage{float}
\usepackage{rotating}

\begin{document}
\title {The role of the $\Lambda(1405)$ in the $pp\rightarrow p K^+\Lambda(1405)$ reaction}

\author{L. S. Geng}\thanks{E-mail address: lsgeng@ific.uv.es}
 \author{E. Oset}\thanks{E-mail address: oset@ific.uv.es}
\affiliation{ Departamento de F\'{\i}sica Te\'orica and IFIC, Centro
Mixto, Institutos de Investigaci\'on de Paterna - Universidad de
Valencia-CSIC, Valencia, Spain}

\begin{abstract}
 We report a theoretical study of the $pp\rightarrow p
K^+\Lambda(1405)$ reaction, which was recently investigated at
COSY-J\"{u}lich by using a 3.65 GeV/c circulating proton beam
incident on an internal hydrogen target. The reaction is driven by
single kaon exchange, single pion exchange, and single rho exchange
terms which have very different shapes due to the two pole structure
of the $\Lambda(1405)$ and the presence of background terms. The
shape for the sum of the three contributions, as well as the total
cross section, are consistent with present data within experimental
and theoretical uncertainties, using reasonable form factors for the meson-baryon vertices.
\end{abstract}
 \maketitle

\section{Introduction}\label{sec:introduction}
The $\Lambda(1405)$ has been a rather controversial resonance for a
long time. In most quark-model calculations, it is described as a
$p$-state $q^3$ baryon with mainly a SU(3) singlet
structure~\cite{Isgur:1978xj}. On the other hand, in
Refs.~\cite{Dalitz:1959dn,Dalitz:1960du}, the $\Lambda(1405)$ is
believed to be a resonance emerging from the interaction of the
$\bar{K}N$ and $\pi\Sigma$ systems, and therefore of $q^4\bar{q}$
structure. Recent studies based on unitary chiral theory,
U$\chi$PT~\cite{Kaiser:1996js,Oset:1997it,Oset:2001cn,Oller:2000fj,Jido:2003cb,Garcia-Recio:2002td,Garcia-Recio:2003ks,Hyodo:2002pk},
in particular favor this interpretation. Furthermore, the models
based on U$\chi$PT  predict that the nominal $\Lambda(1405)$ is a
superposition of two resonances: one  around $1390-i66$\,MeV and the
other around $1426-i16$\,MeV~\cite{Oller:2000fj,Jido:2003cb}. More
recently, the studies of the $\bar{K}N$ interaction  have been
extended by including higher order chiral Lagrangians in the kernel
of the
interaction~\cite{Borasoy:2005ie,Oller:2005ig,Oller:2006jw,Borasoy:2006sr}.
The position of the high-energy pole is rather similar in all these
works, but there are variations in the position of the low-energy
pole. Nevertheless, the theoretical uncertainties have been studied
in \cite{Borasoy:2006sr} and the results obtained with the
lowest-order chiral Lagrangians \cite{Jido:2003cb,Oller:2000fj} are
well within the uncertainties of these extended
models~\cite{Borasoy:2005ie,Oller:2005ig,Oller:2006jw,Borasoy:2006sr}.

As first demonstrated in Ref.~\cite{Jido:2003cb}, due to the fact
that the two poles of the $\Lambda(1405)$ couple differently to the
$\bar{K}N$ and $\pi\Sigma$ channels (the high-energy pole couples
more to the $\bar{K}$N channel whereas the low-energy pole more to
the $\pi\Sigma$ channel),  different production mechanisms may favor
one channel or the other and lead to different invariant mass
distributions, thus offering the possibility to experimentally test
the two-pole prediction. The reactions $\gamma p\rightarrow
K^+\Lambda(1405)$ and $K^-p\rightarrow \Lambda(1405)\gamma$,
(particularly the latter one), are shown to be sensitive to the
high-energy pole of the $\Lambda(1405)$ and thus the corresponding
invariant mass distributions exhibit a peak at $\sim
1420$\,MeV~\cite{Nacher:1998mi,Nacher:1999ni}. On the other hand,
the reaction $\pi^-p\rightarrow K^0(\Sigma\pi)^0$ seems to give more
weight to the low-energy pole and thus exhibits a peak around
$1390$\,MeV in the $\pi\Sigma$ invariant mass
distributions~\cite{Hyodo:2003jw}. Recently, we have shown that the
two-pole structure may also lead to quite different radiative decay
widths~\cite{Geng:2007hz}.

The two-pole structure of the $\Lambda(1405)$ has  inspired several
experimental studies~\cite{Ahn:2003mv}. The Crystal Ball
Collaboration has measured the reaction
$K^-p\rightarrow\pi^0\pi^0\Sigma^0$~\cite{Prakhov:2004an}. In
Ref.~\cite{Magas:2005vu} it was shown that the measured invariant
mass distribution supports the two-pole structure of the
$\Lambda(1405)$.

I. Zychor et al. have recently studied the reaction $pp\rightarrow
pK^+\Lambda(1405)$ at COSY-J\"{u}lich by using a 3.65 GeV/c
circulating proton beam on an internal hydrogen
target~\cite{Zychor:2007gf}. By means of invariant- and missing-mass
techniques, they were able to separate the overlapping
$\Sigma^0(1385)$ and $\Lambda(1405)$. The shape and position of the
$\Lambda(1405)$ constructed from its $\pi^0\Sigma^0$ decay channel
are claimed to be consistent with the data from the
$\pi^-p\rightarrow K^0(\Sigma\pi)^0$ reaction~\cite{Thomas:1973uh}
and the $K^- p\rightarrow \pi^+\pi^-\Sigma^+\pi^-$
reaction~\cite{Hemingway:1984pz}.

It is the main purpose of this paper to study theoretically the
$pp\rightarrow pK^+\Lambda(1405)$ reaction. In sect.~\ref{sec:cc} we
give a brief description of unitary chiral theory and the two
$\Lambda(1405)$'s. In sect.~\ref{sec:reaction} we investigate
possible reaction mechanisms, and build a model based on unitary
chiral theory to study the reaction $pp\rightarrow
pK^+\Lambda(1405)$. In sect.~\ref{sect:results}, we compare the
calculated invariant mass distribution with the data and we show
that our model reproduces rather well both the total cross section
and the invariant mass distribution within the experimental
uncertainties. Summary and conclusions are given in
sect.~\ref{sec:summary}.

\section{Unitary chiral theory and the two $\Lambda(1405)$'s\label{sec:cc}}
 In \cite{Oset:1997it,Oset:2001cn,Oller:2000fj,Jido:2003cb},
 the unitary formalism with coupled channels
 using chiral Lagrangians is exposed. The lowest order chiral
 Lagrangian for the interaction of the pseudoscalar mesons of the SU(3)
 octet of the pion with the baryons of the proton octet is used. By
 picking the terms that contribute to the $MB\rightarrow MB$
 amplitude the Lagrangian is given by \cite{Oset:1997it}:
\begin{equation}
\mathcal{L}=\frac{1}{4f^2}\langle\bar{B}i\gamma^\mu[\Phi\partial_\mu\Phi-\partial_\mu\Phi
\Phi,B]\rangle,
\end{equation}
which, after projected over $s$-wave, provides tree level transition
amplitudes:
\begin{eqnarray}
V_{ij}&=&-C_{ij}\frac{1}{4f^2}(k^0+k'^0),
\end{eqnarray}
with $k^0$($k'^0$) the energy of the initial(final) meson and
$C_{ij}$ are the coefficients tabulated in \cite{Oset:1997it}. These
tree level amplitudes are used as kernel of the Bethe-Salpeter
equation in coupled channels
\begin{equation}
T=[1-VG]^{-1}V,
\end{equation}
where $V$ appears factorized on shell
\cite{Oset:1997it,Oller:2000fj} and $G$ is the loop function of a
meson and a baryon:
\begin{eqnarray}\label{eq:loopf}
G&=&i\int\frac{d^4q}{(2\pi)^4}\frac{M}{E(\vec{q})}\frac{1}{\sqrt{s}-q^0-E(\vec{q})+i\epsilon}\frac{1}{q^2-m^2+i\epsilon}\nonumber\\
&=&\int\frac{d^3q}{(2\pi)^3}\frac{1}{2\omega(\vec{q})}\frac{M}{E(\vec{q})}\frac{1}{\sqrt{s}-\omega(\vec{q})-E(\vec{q})+i\epsilon},
\end{eqnarray}
 which is regularized by a cut off in \cite{Oset:1997it} and in dimensional
regularization in \cite{Oller:2000fj,Oset:2001cn,Jido:2003cb}.

For the particular case of ${1/2}^-$ states (in $MB$ $s$-wave
interaction) with strangeness $S=-1$ and zero charge one has ten
coupled channels: $K^-p$, $\bar{K}^0n$, $\pi^0\Lambda$,
$\pi^0\Sigma^0$, $\eta\Lambda$, $\eta\Sigma^0$, $\pi^+\Sigma^-$,
$\pi^-\Sigma^+$, $K^+\Xi^-$, and $K^0\Xi^0$. The explicit solution
of the Bethe-Salpeter equation leads to poles in the second Riemann
sheet corresponding to resonances. In this sector one finds two
poles close to the nominal $\Lambda(1405)$ resonance, and other
poles corresponding to the $\Lambda(1670)$- and other $\Sigma$-
resonances~\cite{Oset:2001cn,Jido:2003cb}.

In Ref.~\cite{Jido:2003cb}, it was shown that the SU(3)
decomposition of the $1/2^+$ baryon octet and the pseudoscalar meson
octet leads to a singlet and two octets, apart from the 10,
$\overline{10}$, and 27 representations. The two octets are
degenerate in the limit of exact SU(3) symmetry, but the use of the
physical meson and baryon masses breaks explicitly the SU(3)
symmetry and thus the degeneracy. As a result, two branches for
$I=0$ and two for $I=1$ emerge. One of the $I=0$ branch moves to low
energies and comes closer to the singlet at the region of the
nominal $\Lambda(1405)$. Reactions occurring in this region, thus,
would involve both resonances but only an apparent bump would be
seen, giving the impression that there is only one resonance.
However, thanks to the very different couplings of the two poles to
the $\bar{K}N$ and $\pi\Sigma$ channels, and also since the
low-energy pole is broader than the high-energy pole, the shape of
the bump seen is likely to change from one reaction to another.
\begin{table*}[htpb]
\renewcommand{\arraystretch}{1.5}
\setlength{\tabcolsep}{3mm}
 \centering \caption{The $N^*\rightarrow K^+ MB$ couplings with $MB$ one of the ten coupled channels.
 \label{NstarpipiN} }
\begin{tabular}{c|cccccccccc}
\hline\hline
 MB & $K^-p$ & $\bar{K}^0n$ & $\pi^0\Lambda$ & $\pi^0\Sigma^0$ &
 $\eta\Lambda$ & $\eta\Sigma^0$ & $\pi^+\Sigma^-$ & $\pi^-\Sigma^+$
 & $K^+\Xi^-$ & $K^0\Xi^0$\\\hline
 C & 2 & 1 & $\frac{\sqrt{3}}{2}$ & $\frac{1}{2}$ & $\frac{3}{2}$ &
  $\frac{\sqrt{3}}{2}$ & 0 & 1 & 0 & 0\\
 \hline\hline
\end{tabular}
\end{table*}
\begin{figure}[htpb]
 \centering
\includegraphics[scale=0.35]{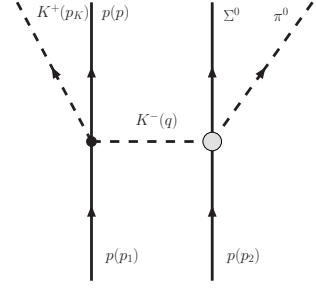}
\caption{\label{kaon_exchange} The kaon exchange mechanism of the
$pp\rightarrow pK^+\Lambda(1405)$ reaction.}
\end{figure}
\begin{figure}[htpb]
 \centering
\includegraphics[scale=0.35]{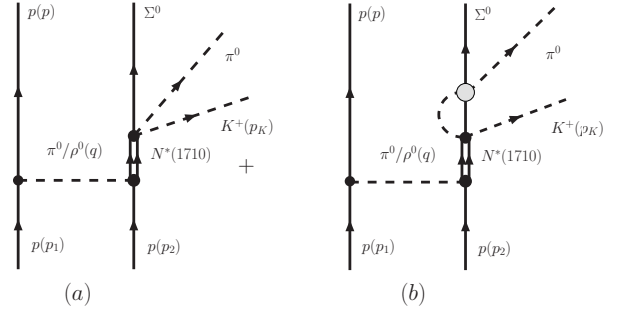}
\caption{\label{pion_exchange} The pion (rho) exchange mechanism of
the $pp\rightarrow pK^+\Lambda(1405)$ reaction through $N^*$
excitation.}
\end{figure}
\begin{figure}[htpb]
 \centering
\includegraphics[scale=0.35]{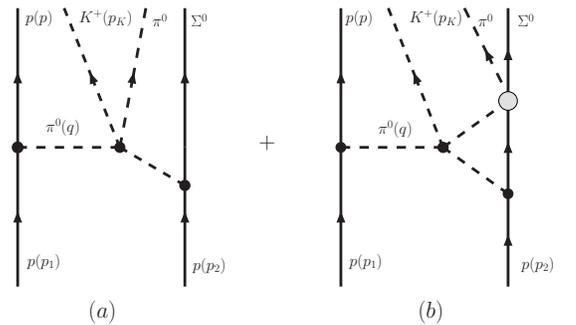}
\caption{\label{meson_pole} The pion exchange mechanism of the
$pp\rightarrow pK^+\Lambda(1405)$ reaction through meson cloud.}
\end{figure}

\section{The $pp\rightarrow p K^+ \Lambda(1405)$ reaction mechanisms}\label{sec:reaction}
In this section, we investigate the possible $pp\rightarrow p K^+
\Lambda(1405)$ reaction mechanisms. We concentrate on the final
decay state $\pi^0\Sigma^0$ of the $\Lambda(1405)$ in order to
compare with the experimental results of \cite{Zychor:2007gf}.
Assuming an $s$-wave for the final states, which are close to
threshold, conservation of spin and parity dictates that the initial
proton-proton system, with isospin $I=1$, has total angular momentum
$L=0$ and total spin $S=0$; therefore, the $pp$ spin wave function
can be written as
\begin{equation}
|pp\rangle=\frac{1}{\sqrt{2}}(|1/2,-1/2\rangle-|-1/2,1/2\rangle).
\end{equation}

With incident protons of laboratory momentum 3.65 GeV/c, the
$pp\rightarrow p K^+ \Lambda(1405)$ reaction can occur through kaon,
pion, and rho meson exchanges, as shown in
Figs.~\ref{kaon_exchange}, ~\ref{pion_exchange}, and
\ref{meson_pole}.

Kaon exchange at low energies for the reaction proceeds as depicted
in Fig.~\ref{kaon_exchange}, where on one nucleon one has the $K^-
p\rightarrow K^-p$ amplitude in $s$-wave, while on the other nucleon
one has the $\Lambda(1405)$ production via
$K^-p\rightarrow\Lambda(1405)\rightarrow\pi^0\Sigma^0$.  With the
strong vertices given in the appendix,  the corresponding $t$ matrix
element reads:
\begin{eqnarray}
t_K=-\frac{1}{2f_K^2}(q^0-\omega_{K^+})\frac{1}{q^{02}-\vec{q}^2-m_{K}^2}t_{K^-p\rightarrow\pi^0\Sigma^0}.
\end{eqnarray}

For the pion exchange mechanism we would have the Yukawa vertex on
one nucleon while the $\pi^0 p\rightarrow
K^+\Lambda(1405)\rightarrow K^+\pi^0\Sigma^0$ reaction on the other
nucleon. For this latter amplitude we follow the method of
Ref.~\cite{Hyodo:2003jw} and we have the mechanisms shown in
Figs.~\ref{pion_exchange} and \ref{meson_pole}, where the pion
excites the $N^*(1710)$ resonance, Fig.~\ref{pion_exchange}, or
interacts with a meson cloud, Fig.~\ref{meson_pole}. In
Ref.~\cite{Hyodo:2003jw} in this latter case the meson pole
amplitude was accompanied with the contact term, and cancellations
between the off shell part of the meson pole term and the contact
term were pointed out there. Here we make explicit use of this
finding and we evaluate only the meson pole term, by taking the
meson meson amplitude on shell (i.e., as a function of the meson
meson invariant mass and replacing $p^2_i$ by $m^2_i$ for the meson
legs).

The $t$ matrix element corresponding to pion exchange through $N^*$
excitation (Fig.~\ref{pion_exchange}) reads
\begin{eqnarray}\label{eq:tpi}
t_\pi&=&-\frac{g_A}{2f_\pi}\vec{q}^2\frac{1}{q^{02}-\vec{q}^2-m^2_\pi}\frac{1}{\sqrt{s'}-M_{N^*}+i\frac{\Gamma}{2}}
\frac{AB}{f_\pi^3}\times\\
&&\left[C_{\pi^0\Sigma^0}(\omega_{\pi^0}-\omega_{K^+})+ \sum_i
C_i(\omega_i-\omega_{K^+})G_i
t_{i\rightarrow\pi^0\Sigma^0}\right],\nonumber
\end{eqnarray}
where $s'$ is the  invariant energy squared of the $N^*(1710)$
determined by
\begin{equation}
s'=(p_1+p_2-p)^2=s+M^2_N-2\sqrt{s}E(p),
\end{equation}
$G_i$ are the one-meson one-baryon loop functions of
Eq.~(\ref{eq:loopf}), $C_i$ the coupling constants tabulated in
Table \ref{NstarpipiN}, and $i$ one of the ten coupled channels. $A$ and $B$ are the
coupling constants of the $N^*(1710)$ decaying into baryon-meson and baryon-meson-meson as defined
in the appendix and in Ref.~\cite{Hyodo:2003jw}.
Their numerical values are given below. The
amplitudes $t_{MB\rightarrow MB}$ are the meson-baryon meson-baryon
amplitudes described in sec.~\ref{sec:cc}. Finally, the $t$-matrix
element corresponding to pion exchange through meson cloud
(Fig.~\ref{meson_pole}) reads
\begin{eqnarray}\label{eq:mp}
t_\mathrm{MP}=\frac{g_A}{2f_\pi}\vec{q}^2\frac{1}{q^{02}-\vec{q}^2-m^2_\pi}
\left[\mathcal{M}_{4}+ \sum_{i} \mathcal{M}_iG_i t_{i\rightarrow
4}\right]
\end{eqnarray}
with the amplitudes $\mathcal{M}_i$ given in the appendix.

 The $N^*(1710)$ decay coupling
constants $A$ and $B$ appearing in Eq.~(\ref{eq:tpi}) are fixed by
the $N^*(1710)$ partial decay widths into $\pi\pi N$ and  $\pi N$.
Considering the rather large uncertainty of the $N^*(1710)$ total
decay width and the corresponding branching
ratios~\cite{Yao:2006px}, we choose two sets of parameters: For
parameter set I, we take $M_{N^*}=1740$\,MeV, $\Gamma=200$\,MeV,
$\Gamma_{\pi\pi N}=100$\,MeV, $\Gamma_{\pi N}=$40\,MeV, which yield $A=0.11$ and
$B=0.84$; for
parameter set II, we take $M_{N^*}=1710$\,MeV, $\Gamma=100$\,MeV,
$\Gamma_{\pi\pi N}=65$\,MeV, $\Gamma_{\pi N}$=15\,MeV, which yield $A=0.07$ and $B=0.77$. A recent
combined analysis~\cite{Sarantsev:2007bk} of different reactions
also require the presence of the $N^*(1710)$ but it has not improved on
the present uncertainties of the properties of this resonance. For the energy
evolution of the $N^*$ decay width, we have taken into account the
effects of the Blatt-Weisskopf penetration
factors~\cite{Manley:1984jz,Manley:1992yb}.

The $N^*$ excitation mechanism can also be induced by $\omega$ or
$\rho$ meson exchange (see Fig.~\ref{pion_exchange}). We note that
the latest study of \cite{Muehlich:2006nn} shows that the branching
ratio of the $N^*(1710)$ decaying into $N\omega$ is only $0.2\%$,
instead of $(13.0\pm2.0)\%$, as quoted by the PDG~\cite{Yao:2006px},
deduced by the same authors as in \cite{Muehlich:2006nn} in an
earlier work~\cite{Penner:2002ma}. Therefore, we will not consider
the $N^*$ excitation induced by an $\omega$ meson. According to the
PDG~\cite{Yao:2006px}, the branching ratio of the $N^*(1710)$
decaying into $N\rho$ is $5-25\%$.  With the standard strong
vertices as shown in the appendix, the $t$-matrix element of the
$\rho$ induced $N^*$ excitation reads
\begin{equation}
t_\rho=t_\rho^{(1)}+t_\rho^{(2)}\frac{(\sigma^{(1)}\times
\vec{q})(\sigma^{(2)}\times \vec{q})}{\vec{q}^2}
\end{equation}
with
\begin{eqnarray*}
t_\rho^{(1)}&=&-\mathcal{N}^2\Bigg\{\left[G^V-\frac{\vec{q}^2}{2M(E+M)}G^T\right]\left(1-\frac{q^{02}}{m^2_\rho}\right)\nonumber\\
&&+\frac{G^V
\vec{q}^2}{(E+M)^2}\left(1+\frac{\vec{q}^2}{m_\rho^2}\right)+
\frac{G^T}{2M(E+M)^2}\frac{q^0\vec{q}^4}{m_\rho^2}\Bigg\}\nonumber\\
&&\times \frac{1}{q^{02}-\vec{q}^2-m^2_\rho}G_{\rho
NN^*}\frac{1}{\sqrt{s'}-M_{N^*}+i\frac{\Gamma}{2}}
\frac{B}{f_\pi^2}\times\\
&&\left[C_{\pi^0\Sigma^0}(\omega_{\pi^0}-\omega_{K^+})+ \sum_i
C_i(\omega_i-\omega_{K^+})G_i t_{i\rightarrow\pi^0\Sigma^0}\right],
\end{eqnarray*}
\begin{eqnarray*}
t^{(2)}_\rho&=&\mathcal{N}^2\frac{\vec{q}^2}{E+M}\left(\frac{G^V}{E+M}+\frac{G^T}{2M}\right)\nonumber\\
&&\times
\frac{1}{q^{02}-\vec{q}^2-m^2_\rho}G_{\rho
NN^*}\frac{1}{\sqrt{s'}-M_{N^*}+i\frac{\Gamma}{2}}
\frac{B}{f_\pi^2}\times\\
&&\left[C_{\pi^0\Sigma^0}(\omega_{\pi^0}-\omega_{K^+})+ \sum_i
C_i(\omega_i-\omega_{K^+})G_i
t_{i\rightarrow\pi^0\Sigma^0}\right],\nonumber
\end{eqnarray*}
where $\mathcal{N}=\sqrt{\frac{E+M}{2M}}$; $E$ and $M$ are the
energy and mass of the initial protons.

The value of the coupling constant $G_{\rho NN^*}$ in the $G_{\rho
NN^*}\gamma^\mu\epsilon_\mu\vec{\tau}\cdot\vec{\rho}$ vertex is
determined by reproducing the $N^*(1710)$ decay width into $N\rho$
($\sim15$\,MeV). Its sign, however, cannot be fixed. Therefore, we
will present results corresponding to both cases. Taking into
account the relatively large widths of the $N^*(1710)$ and the rho,
we used the following double convolution to obtain the $N^*(1710)$
decay width into $N\rho$:
\begin{eqnarray}
\Gamma_\rho&=&\frac{3}{C}\int\limits^{M_{N^*}+2\Gamma}_{M_{N^*}-2\Gamma} d\tilde{M}\int\limits^{m_\rho+2\Gamma_\rho}_{m_\rho-2\Gamma_\rho}
d\tilde{m_\rho}\\
&&\times\frac{1}{\pi}\mathrm{Im}\frac{1}{\tilde{M}-M_{N^*}+i\Gamma/2}
G_{\rho NN^*}^2\nonumber\\
&&\times
\frac{1}{\pi}\mathrm{Im}\frac{1}{\tilde{m}_\rho-m_\rho+i\Gamma_\rho/2}
\Theta(\tilde{M}-\tilde{m}_\rho-M)\nonumber\\
&&\times\frac{1}{2\pi}\frac{M}{\tilde{M}}q\left(-\frac{3}{2}+\frac{E-q^0}{2M}
+\frac{q^0(\tilde{M}^2-M^2)}{2M \tilde{m}_\rho^2}\right),\nonumber
\end{eqnarray}
where $\Theta$ is the step function, $q$ is the 3-momentum of
$N\rho$ in the rest frame of the $N^*(1710)$, $E$ and $M$ the
nucleon energy and mass, the factor 3 accounts for isospin, and $C$
is the normalization constant
\begin{eqnarray}
C&=&\int\limits^{M_{N^*}+2\Gamma}_{M_{N^*}-2\Gamma} d\tilde{M}\int\limits^{m_\rho+2\Gamma_\rho}_{m_\rho-2\Gamma_\rho}
d\tilde{m_\rho}\frac{1}{\pi}\mathrm{Im}\frac{1}{\tilde{M}-M_{N^*}+i\Gamma/2}\nonumber\\
&&\hspace{0.2cm}\times
\frac{1}{\pi}\mathrm{Im}\frac{1}{\tilde{m}_\rho-m_\rho+i\Gamma_\rho/2}\Theta(\tilde{M}-\tilde{m}_\rho-M).
\end{eqnarray}
For the mass and width of the $N^*(1710)$ appearing in the above
equation, we use the $N^*$ parameter set I.

 With incident protons of lab
momentum 3.65 GeV/c, the exchanged pion and rho are very much off
shell, which must be taken into account in a realistic study. For
the pion exchange diagram, we multiply the $\pi NN$ vertex by the
following recoil correction
 \begin{equation}
 R(q)=1-\frac{q^0}{2M}
 \end{equation}
 with $q$ outgoing from the nucleon and the form factor
 \begin{equation}
 F(q)=\frac{\Lambda_\pi^2-m^2_\pi}{\Lambda_\pi^2-q^2}
 \end{equation}
 with $\Lambda_\pi=1.0$\,GeV. For the $\pi N N^*$ vertex, we
 multiply the same form factor but with the recoil correction:
 \begin{equation}
 R(q)=1+\frac{q^0}{2M},
 \end{equation}
 since $q$ is now incoming. For the $\rho NN$ and $\rho N N^* $ vertices, following
 Ref.~\cite{Machleidt:1987hj}, we multiply a form factor of the
 dipole form
 \begin{equation}
 F(q)=\left(\frac{\Lambda_\rho^2-m^2_\rho}{\Lambda_\rho^2-q^2}\right)^2
 \end{equation}
 with $\Lambda_\rho=2$\,GeV. As for the kaon exchange diagram,
 taking into account relativistic correction, the $p\rightarrow p KK$
 vertex  becomes
 \begin{equation}
-it=i\frac{1}{2f_K^2}(\omega_{K^-}-\omega_{K^+}-E_N+M_N).
\end{equation}
We also multiply this vertex with a form factor of the  form
\begin{equation}\label{kaonff}
 F(q)=\frac{\Lambda^2_K-m^2_K}{\Lambda^2_K-q^2}
\end{equation}
with $\Lambda_K=1.25$ GeV.  A moderate modification of all the cutoff values, $\Lambda_\pi$, $\Lambda_\rho$, and $\Lambda_K$,
will not change our results significantly, as will be shown below.

\section{Results and discussion}\label{sect:results}

With all the $t$-matrix elements provided above, the invariant mass
distribution is then calculated by
 \begin{eqnarray}\label{eq:inv}
 \frac{d\sigma}{dM_{\pi^0\Sigma^0}}&=&\frac{1}{32\pi^5}\frac{M^3_N
 M_{\Sigma^0}}{\sqrt{s^2-4s\,M_N^2}}\int dE\int
 d\omega\,\Theta(1-\cos^2\theta)\nonumber\\
 &&\hspace{0.5cm}\times\tilde{k}\bigg\{
 [2(t_K+t^{(1)}_\rho)]^2+2[2t^{(2)}_\rho]^2\nonumber\\
 &&\hspace{3.5cm}+[2(t_\pi+t_\mathrm{MP})]^2\bigg\},
 \end{eqnarray}
with $\tilde{k}$ the $\pi^0$ ($\Sigma^0$) 3-momentum in the
center-of-mass frame of $\pi^0\Sigma^0$
 and $\theta$ the angle between $\vec{p}$ and $\vec{p}_K$
 given by
\begin{equation}
\tilde{k}=\frac{\lambda^{1/2}(M^2_{\pi^0\Sigma^0},m^2_{\pi^0},M^2_{\Sigma^0})}{2M_{\pi^0\Sigma^0}},
\end{equation}
\begin{equation}
\cos\theta=\frac{(\sqrt{s}-E-\omega)^2-\vec{p}^2-\vec{p}_K^2-M^2_{\pi^0\Sigma^0}}{2|\vec{p}||\vec{p}_K|},
\end{equation}
where $E$ and $\omega$ are the energies of the final proton and
$K^+$. The factors 2 in Eq.~(\ref{eq:inv}) accounts for the
possibility of having the $\Lambda(1405)$ production from either of
the two protons.
\begin{figure}[b]
 \centering
\includegraphics[scale=0.34,angle=270]{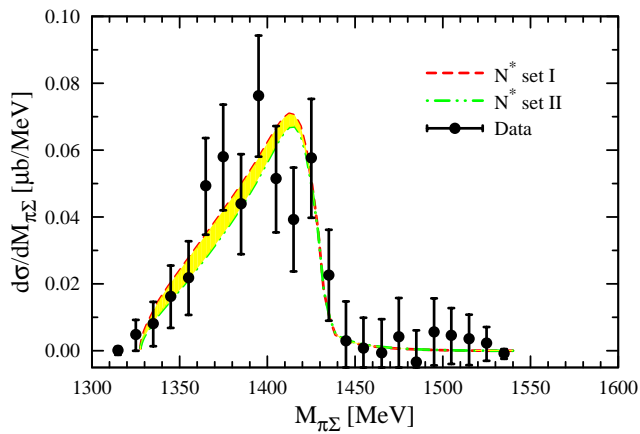}
\caption{\label{fig_inv1} The invariant mass distribution of the
$\pi\Sigma$ in comparison with the data~\cite{Zychor:2007gf}.
The $\rho$ exchange contribution is not included.}
\end{figure}
\begin{figure}[t]
 \centering
\includegraphics[scale=0.34,angle=270]{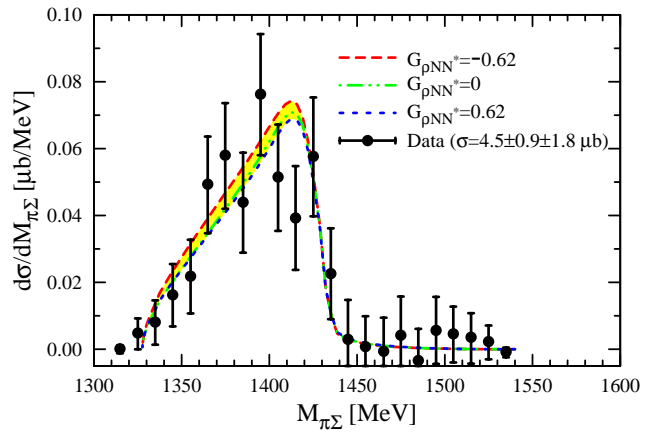}
\caption{\label{fig_inv} The invariant mass distribution of the
$\pi\Sigma$ in comparison with the data~\cite{Zychor:2007gf}.}
\end{figure}

It is to be noted that although Ref.~~\cite{Zychor:2007gf}
only measured the $\pi^0\Sigma^0$ final state,
the total cross section given in Ref.~\cite{Zychor:2007gf} is for $\Lambda(1405)\rightarrow
\pi\Sigma$, which implies that a factor of 3 has been multiplied to account for the isospin. To compare with the data,
we have multiplied our invariant mass distribution , Eq.~(\ref{eq:inv}),
with the same factor to obtain the distribution of the $\pi\Sigma$. 

In Fig.~\ref{fig_inv1}, the calculated invariant mass distribution
of the $\Lambda(1405)$ with $N^*$ parameter sets I and II are
compared with the new data of Ref.~\cite{Zychor:2007gf}.
The shaded
area indicates the uncertainties of our calculation related to the
determination of the $N^*$ coupling constants $A$ and $B$. For
demonstration purposes, we did not include the $\rho$ exchange
contribution. It is seen that within the experimental uncertainties,
our calculations reproduce the data rather well, particularly the
fast drop at the $\bar{K}N$ threshold. Although both parameter sets
reproduce the data very well, in particular taking into account the
large experimental uncertainties, we would say that parameter set I
is preferred, which is in agreement with the finding of
Ref.~\cite{Hyodo:2003jw}. In the following, we would use parameter
set I as our default choice.

\begin{figure}[t]
 \centering
\includegraphics[scale=0.34,angle=270]{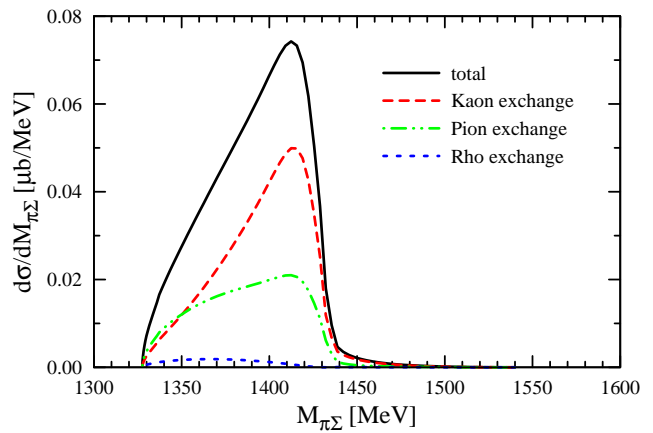}
\caption{\label{fig_component} The contribution of the three
different mechanisms with $G_{\rho N N^*}=-0.62$.}
\end{figure}
\begin{figure}[t]
 \centering
\includegraphics[scale=0.34,angle=270]{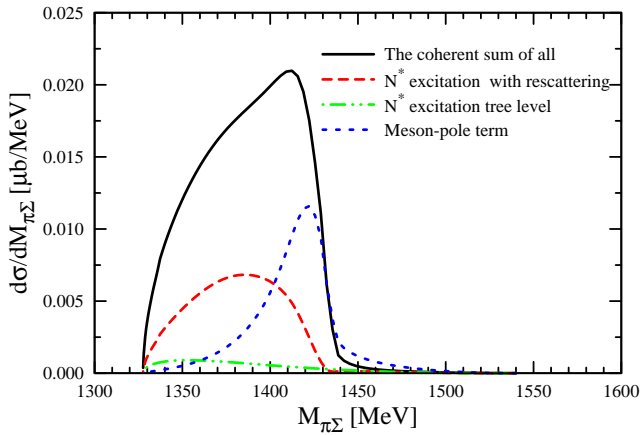}
\caption{\label{fig:tpi_inpart} The components of the $\pi$ exchange
contribution. See text for details.}
\end{figure}
Now we would like to study the contribution of the $\rho$ exchange.
The coupling constant $G_{\rho NN^*}$ is fixed to reproduce the
estimated $N^*(1710)$ decay width into $N\rho$, $\sim15$\,MeV, which
yields $|G_{\rho N N^*}|=0.62$. Its sign, however, cannot be fixed.
In Fig.~\ref{fig_inv}, we present the calculated invariant mass
distribution corresponding to both cases, i.e. $G_{\rho NN^*}=-0.62$
and $G_{\rho NN^*}=0.62$. It is seen that both reproduce the data
rather well, in other words, the quality of the present data cannot
discriminate the sign of $G_{\rho NN^*}$. We further notice that our
calculated total cross section $\sim5$ $\mu b$ is also in good
agreement with the data $4.5\pm0.9\pm1.8$ $\mu b$.

In Fig.~\ref{fig_component}, the contribution of the kaon exchange
mechanism and those of the pion and rho exchanges are compared. 
It can be clearly seen that the kaon exchange mechanism leads to 
an asymmetric peak at $\sim1410$\,MeV, while the pion exchange
mechanism broadens the shape and leads to a better
agreement with the data. We would like to stress that the $\rho$
exchange contribution by itself is very small, only through the
interference with the kaon contribution its effect becomes
relevant.

The broad shape of the pion exchange mechanism is actually made by
the collaboration of three very different contributions as can be
seen in Fig.~\ref{fig:tpi_inpart}. One comes from the tree level
diagram (diagram (a) of Fig.~\ref{pion_exchange}), which peaks at
low invariant masses. Another one is from the mechanism with re-scattering (diagram (b) of
Fig.~\ref{pion_exchange}), which is dominated by the broad
$\Lambda(1405)$ pole of low energy. Finally the mechanism of the
meson pole, Fig.~\ref{meson_pole}, is dominated by the narrow high
energy pole of the $\Lambda(1405)$. The coherent sum of all these
mechanisms produces the broad shape shown in
Fig.~\ref{fig_component}. One can see in this figure that the pion
exchange term provides strength for the $pp\rightarrow
K^+\pi^0\Sigma^0$ reaction in the low energy side of the invariant
mass, leading to an apparent broader width of the $\Lambda(1405)$
compared with the one we would obtain from the $K$ exchange
mechanism alone, which is mostly dominated by the high energy
$\Lambda(1405)$ pole.

It is interesting to note that the strong amplitudes
$t_{MB\rightarrow MB}$ are determined by the very precise $\bar{K}N$
branching ratios $r$, $R_c$, and $R_n$~\cite{Oset:1997it};
therefore,  most uncertainties in our model come from the
$N^*(1710)$ coupling constants $A$, $B$, and $G_{\rho N N^*}$, which
are partly shown in Figs.~\ref{fig_inv1} and ~\ref{fig_inv}, and the
form factors which take into account the off-shellness of the
exchanged particles.

\begin{figure}[t]
 \centering
\includegraphics[scale=0.34,angle=270]{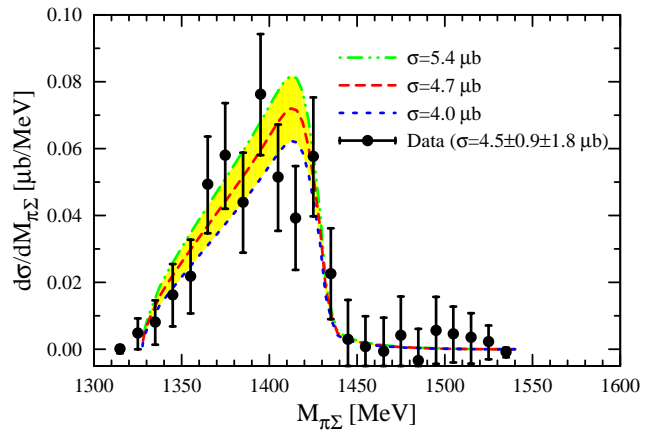}
\caption{\label{fig:error} The invariant mass distribution of the
$\pi\Sigma$ with theoretical uncertainties estimated using
Monte-Carlo sampling method (see text for details), in comparison
with the data~\cite{Zychor:2007gf}.}
\end{figure}

To estimate the theoretical uncertainties related to the $N^*(1710)$
decay widths and the cutoff values, we perform a Monte-Carlo
sampling of the parameter values within their uncertainties, more
specifically, we allow the $N^*(1710)$ total width to vary in the
range of $200\pm30$\,MeV, $\Gamma_{\pi\pi N}$ in the range of
$100\pm15$\,MeV, $\Gamma_{\pi N}$ in the range of $40\pm6$\,MeV,
$\Gamma_{\rho N}$ in the range of 5-25\,MeV, $\Lambda_\pi$ within
$1.0\pm 0.1$\,GeV, $\Lambda_\rho$ within $2.0\pm 0.2$\,GeV, while $\Lambda_K$ within $1.25\pm0.125$\,GeV.
Since $G_{\rho N N^*}$ can have either negative or positive sign, we
assign half of the Monte-Carlo generated values positive sign and
half of them negative sign. The so-obtained averaged
invariant mass distribution and the band corresponding to
$\bar{y}\pm\sigma$ with $\bar{y}=\frac{1}{N}\sum y_i$ and
$\sigma^2=\frac{1}{N-1}\sum(y_i-\bar{y})^2$ are displayed in
Fig.~\ref{fig:error}. The total cross section is estimated to be
$4.7\pm 0.7\,\mu\mathrm{b}$, which should be compared with the
data: $4.5\pm0.9\pm1.8$ $\mu b$~\cite{Zychor:2007gf}.

The strength of the present reaction has brought a new information concerning the kaon exchange diagram in 
Fig.~\ref{kaon_exchange}, where we implemented a form factor in the $pK\bar{K}$ vertex. Should we have not taken this form factor 
into account, the contribution of the kaon exchange, which is the dominant mechanism, would have been larger and the cross section would have been about a
factor of three times bigger than what we have evaluated. We can state this, but cannot be more conclusive with respect to the shape
of the form factor, Eq.~(\ref{kaonff}), because the present experiment selects only one value of $q^2$ approximately, $q^2\approx-|\vec{p}_1|^2\approx
-(1150\,\mbox{MeV})^2$. Similarly, we cannot induce whether a form factor should be implemented in only one or
 both vertices of the kaon exchange. All that
the experiment is telling us is that for the off-shellness of this process, with $q^2\approx
-(1150\,\mbox{MeV})^2$, the kaon exchange amplitude
 is reduced by about a factor of two with respect to the ordinary kaon exchange
with no form factors.

\section{Summary and conclusions}\label{sec:summary}
We have performed a theoretical study of the $pp\rightarrow
pK^+\Lambda(1405)$ reaction recently investigated at
COSY-J\"{u}lich. Based on unitary chiral theory, we constructed a
model including three different mechanisms: single-kaon exchange,
single-pion exchange, and single-rho exchange. We showed that the
kaon exchange mechanism was mostly sensitive to the high energy pole
of the $\Lambda(1405)$ and produced by itself a relatively narrow
structure. Yet, the mechanism of pion exchange, itself a combination
of various terms, provided a  different shape for the
$\pi^0\Sigma^0$ invariant mass, that added to the one of kaon
exchange had the effect of producing strength in the low invariant
mass part, resulting in a broadening of the invariant mass
distribution and a better agreement with experiment. The rho
exchange contribution has a similar (but smaller) effect as the pion
exchange if the sign of $G_{\rho NN^*}$ is negative. 

The total strength of the cross section demanded a reduction of the dominant kaon exchange mechanism,
where the kaon appears largely off shell. The reduction is of the order of a factor of two in the amplitude for 
$q^2\approx
-(1150\,\mbox{MeV})^2$. Thus we  introduced a monopole form factor
with $\Lambda_K\approx1.25$ GeV, which is of natural size. With this value chosen, an error analysis was performed by
changing
input parameters within experimental boundaries, or changing the cutoff parameters of the form factors by about $10\%$ of their
central values. The result is a band of cross sections with a certain dispersion at low
invariant masses, compatible with experiment, and a fast fall down around 1430 MeV, 
rather independent of the input, which is also clearly seen in the data.

Once more we show in this paper that the association of a shape of
the $\pi^0\Sigma^0$ distribution to a universal $\Lambda(1405)$
resonance is a delicate subject and that one should rather make a
thorough study of the different reaction mechanisms entering the
process, since the observed final shape is a subtle combination of
contributions from background and the two $\Lambda(1405)$ poles
which are weighted differently in the various mechanisms.

\section{Acknowledgments}
We would like to acknowledge useful discussions with I.
Zychor, H. Str\"{o}her, and  particularly the 
lengthy and instructive discussions with C. Wilkin.
  L. S. Geng acknowledges financial
support from the Ministerio de Educacion y Ciencia in the Program of
estancias de doctores y tecnologos extranjeros. This work is partly
supported by DGICYT contract number FIS2006-03438 and the
Generalitat Valenciana. This research is part of the EU Integrated
Infrastructure Initiative Hadron Physics Project under contract
number RII3-CT-2004-506078.

\section*{Appendix}
 For the strong vertices, we have used the following
Feynman rules:
\begin{enumerate}

\item $p\rightarrow p\pi^0$:
\begin{equation}
-it=-\frac{g_A}{2f_\pi}\vec{\sigma}\cdot\vec{q}.
\end{equation}

\item $p\pi^0\rightarrow N^*$:
\begin{equation}
-it=\frac{A}{f_\pi}\vec{\sigma}\cdot\vec{q}.
\end{equation}

\item $ p\rho\rightarrow  N^*$:
\begin{equation}
-it=iG_{\rho N
N^*}\bar{\psi}\gamma^\mu\vec{\tau}\cdot\vec{\epsilon}_\mu\psi.
\end{equation}

\item $p\rightarrow p\rho$:
\begin{equation}
-it=-i\bar{\psi}\left\{G^V\gamma^\mu+G^T\frac{1}{2i
m_p}\sigma^{\mu\nu}
q_\nu\right\}\vec{\tau}\cdot\vec{\epsilon}_\mu\psi,
\end{equation}
with $G^V=2.9\pm 0.3$ and
$G^T/(2m_p)=1.35\pm0.12\,m_\pi^{-1}$, which
provide $G^T/G^V=6.27$~\cite{Oset:1985wt}.

\item $N^*\rightarrow K^+ MB$:
\begin{equation}
-it=i\frac{B}{f_\pi^2}C_{MB}(\omega_M-\omega_{K^+}),
\end{equation}
with the $C_{MB}$ couplings tabulated in Table \ref{NstarpipiN}.

\item $p\rightarrow p K^+ K^-$:
\begin{equation}
-it=i\frac{1}{2f_K^2}(\omega_{K^-}-\omega_{K^+}).
\end{equation}

\item $MM\rightarrow MM$
\begin{equation}
-it=i\frac{1}{12f_\pi^2}\langle(\partial\Phi\Phi-\Phi\partial\Phi)^2+M\Phi^4\rangle,
\end{equation}
with $\Phi$ the meson octet and $M$ the quark mass matrix.

\item $MB\rightarrow B$
\begin{equation}
-it=i\left[\frac{D+F}{2}\langle\bar{B}\gamma^\mu\gamma_5 u_\mu
B\rangle+\frac{D-F}{2}\langle\bar{B}\gamma^\mu\gamma_5 B
u_\mu\rangle\right],
\end{equation}
\end{enumerate}
with $B$ the baryon octet of the proton. In the present work, the
following parameter values have been used: $f_\pi=93$ MeV, $f_K=1.22
f_\pi$, $g_A=1.26$, $D=0.795$, and $F=0.465$.

When calculating the meson-pole diagram, as explained in detail in
Ref.~\cite{Hyodo:2003jw}, one can put the $MMMM$ vertex on shell,
the off shell part after canceling a meson-propagator, will be
canceled by the $MMBBB$ contact term~\cite{Hyodo:2003jw}.
Furthermore, with the present experimental setup, one can assume
that the outgoing mesons and baryons are almost at rest, which we
take for the evaluation of matrix elements. Below, we give the
meson-pole amplitudes appearing in Eq.~(\ref{eq:mp}) in the order of
$K^-p$, $\bar{K}^0n$, $\pi^0\Lambda$, $\pi^0\Sigma^0$,
$\eta\Lambda$, $\eta\Sigma^0$, $\pi^+\Sigma^-$, $\pi^-\Sigma^+$,
$K^+\Xi^-$, and $K^0\Xi^0$:

\begin{equation}
 \mathcal{M}_1=\mathcal{M}^\pi_1+\mathcal{M}^\eta_1,
\end{equation}

\begin{equation}
\mathcal{M}_1^\pi=-\frac{D+F}{6f_\pi^3}\frac{s_1}{D_1(m_\pi)},
\end{equation}
\begin{eqnarray}
\mathcal{M}_1^\eta&=& \frac{D-3F}{24f^3_\pi}\frac{1}{D_1(m_\eta)}\\
&&\times\left[3s_1-\frac{1}{3}m^2_\pi-m_\eta^2-\frac{8}{3}m^2_K\right],\nonumber
\end{eqnarray}
\begin{eqnarray}
\mathcal{M}_3&=&\frac{3F+D}{12\sqrt{3}f^3_\pi}\frac{1}{D_3(m_K)}\\
&&\times\left[-s_3+2(m^2_\pi+m^2_K)-\frac{(m^2_K-m^2_\pi)^2}{s_3}\right],\nonumber
\end{eqnarray}
\begin{eqnarray}
\mathcal{M}_4&=&\frac{F-D}{12f^3_\pi}\frac{1}{D_4(m_K)}\\
&&\times\left[-s_4+2(m^2_\pi+m^2_K)-\frac{(m^2_K-m^2_\pi)^2}{s_4}\right],\nonumber
\end{eqnarray}
\begin{eqnarray}
\mathcal{M}_5&=&\frac{3F+D}{24
f^3_\pi}\frac{1}{D_5(m_K)}\\
&&\times\left[-\frac{3}{2}s_5-\frac{(m^2_\pi-m^2_K)(m^2_\eta-m^2_K)}{2s_5}\right.\nonumber\\
&&\left.\hspace{2.5cm}+\frac{7
}{6}m^2_\pi+\frac{m^2_\eta}{2}+\frac{1}{3}m^2_K\right],\nonumber
\end{eqnarray}
\begin{eqnarray}
\mathcal{M}_6&=&\frac{\sqrt{3}(F-D)}{24
f^3}\frac{1}{D_6(m_K)}\\
&&\times\left[-\frac{3}{2}s_6-\frac{(m^2_\pi-m^2_K)(m^2_\eta-m^2_K)}{2s_6}\right.\nonumber\\
&&\hspace{2.5cm}\left.+\frac{7}{6}m^2_\pi+\frac{m^2_\eta}{2}+\frac{1}{3}m^2_K\right],\nonumber
\end{eqnarray}
\begin{eqnarray}
\mathcal{M}_8&=&\frac{D-F}{4f^3_\pi}\frac{1}{D_8(m_K)}\\
&&\times\left[s_8-m^2_\pi-m^2_K+\frac{s^2_8-(m^2_K-m^2_\pi)^2}{2s_8}\right],\nonumber
\end{eqnarray}
\begin{equation}
\mathcal{M}_2=\mathcal{M}_7=\mathcal{M}_9=\mathcal{M}_{10}=0.
\end{equation}
The meson propagator is given by
\begin{equation}
 \frac{1}{D_i(m)}=\frac{1}{(q^0-\omega_i-m_K)^2-\vec{q}^2-m^2},
 \end{equation}
 with the energy of the meson $\omega_i$ and the invariant mass squared of the meson-pair $s_i$
 given by
 \begin{equation}
 \omega_i=\frac{M^2_I+m^2_i-M^2_i}{2M_I},
 \end{equation}
 \begin{equation}\label{eq:pair_inv}
 s_i=(m_K+M_I-M_i)^2,
 \end{equation}
 where $M_I$ is the invariant mass of the $\Lambda(1405)$ and $m_i$ ($M_i$) the
meson (baryon) mass of channel $i$. The last equation,
Eq.~(\ref{eq:pair_inv}), is obtained assuming final particles with
small momentum, in line with the comments made above, and is
sufficiently good for our purpose.
 \bibliographystyle{apsrev}
 \bibliography{paper}

\end{document}